# A hidden bulk polymorph governs charge transport dimensionality in an organic semiconductor

Caterina Zuffa[1], Marco Bardini[2], Fabian Gasser[3], Mauricio Sevilla[4], Robinson Cortes-Huerto[4], Alessandro Greco[4], Lorenzo Soprani[2,5], Guanzhao Wen[4], Jaco J. Geuchies[4,8], Mischa Bonn[4], Gabriele D'Avino[6,2*], Lucia Maini[1*], Hai I. Wang[4,7*], Lucia Di Virgilio[4*]

[1] Dipartimento di Chimica 'G. Ciamician', University of Bologna, via Gobetti, 85, Bologna 40129, Italy

[2] Grenoble Alpes University, CNRS, Grenoble INP, Institut Néel, Grenoble 38042, France

[3] Institute of Solid State Physics, Graz University of Technology, Petersgasse 16, Graz, 8010, Austria

[4] Max Planck Institute for Polymer Research, Mainz 55128, Germany,

[5]Dipartimento di Chimica Industriale "Toso Montanari", Università di Bologna, Bologna 40129Italy

[6] Department of Molecular Sciences and Nanosystems, Ca' Foscari University of Venice, Venice 30123, Italy

[7] Nanophotonics, Debye Institute for Nanomaterials Science, Utrecht University, Princetonplein 1, Utrecht, The Netherlands

[8] Leiden Institute of Chemistry, Leiden University, Einsteinweg 55, 2333CC, Leiden, the Netherlands

*Correspondence to: divirgiliol@mpip-mainz.mpg.de, h.wang5@uu.nl; l.maini@unibo.it, gabriele.davino@unive.it

**Abstract**

Organic semiconductors (OSCs) are widely explored for flexible optoelectronic technologies, with performance governed not only by molecular design, but also by solid-state packing, which can give rise to polymorphism. Dinaphtho[2,3-b:2′,3′-f]thieno[3,2-b]thiophene (DNTT) is a benchmark OSC that has long been considered monomorphic. Here, we discover, isolate, and resolve the crystal structure of a previously unrecognised bulk polymorph of DNTT, termed ‘blue DNTT’ owing to its characteristic blue emission. Coexisting with the well-known (*green)* DNTT in commercial powders – yet previously overlooked – *blue* DNTT represents the thermodynamically stable form. By combining X-ray diffraction, Raman, and THz spectroscopy with simulations, we demonstrate that polymorphism in DNTT reshapes the low-frequency phonon landscape and transfer-integral network, impacting charge transport. While *green* DNTT exhibits two-dimensional charge transport with holes more mobile than electrons, *blue* DNTT shows charge transport along all crystallographic directions enabled by a distinct herringbone packing. Electron mobility along the crystallographic a- and b-axes in *blue* DNTT exceeds twice the hole mobility in the *green* phase. To our knowledge, this is the first reported acene-based semiconductor exhibiting three-dimensional charge transport. Polymorphism emerges as a key lever to tune charge transport dimensionality and carrier efficiency in organic semiconductors.

## Introduction

Organic semiconductors (OSCs) have attracted significant attention for optoelectronic applications, with several systems exhibiting electrical mobilities higher than 10 $\mathrm{cm}^2/\mathrm{V\,s}$. Among them, oligoacenes stand out for their exceptionally high charge-carrier mobilities, reaching up to 20 $\mathrm{cm}^2/\mathrm{V\,s}$ in rubrene crystals [1]. The extended $\pi$-structure in oligoacenes enables substantial overlap of intermolecular electronic orbitals along two directions, exhibiting two-dimensional (2D) charge transport. As the number of benzene rings increases, the HOMO energy increases and the HOMO-LUMO gap decreases [2]. These effects render molecules susceptible to oxidation under ambient conditions, in particular upon photoexcitation. Therefore, despite their excellent 2D charge-transport properties, oligoacenes suffer from poor chemical and environmental stability, limiting their widespread use in organic solar cells and organic field-effect transistors under standard operating conditions.

In this context, Takimiya and coworkers developed a new strategy to overcome these instability issues by using heteroatoms with large atomic radii, such as sulfur in thiophene rings, into the acene backbone[3]. The replacement of selected benzene rings with thiophene units enhanced chemical stability and optimised intermolecular interactions, leading to the thienoacene family. Using molecular orbital (MO) calculations, they designed the target molecules with an extended $\pi$ conjugated system for high electrical mobility and a low-lying HOMO level to enhance air stability [2]. A notable example is dinaphtho[2,3-b:2′,3′-f]thieno[3,2-b]thiophene (DNTT), consisting of two thiophene rings flanked by two pairs of benzene rings (see **Fig. 1a**). Since its discovery, DNTT has immediately emerged as one of the best-performing OSCs with charge mobility of up to 8-13 $\mathrm{cm}^2/\mathrm{V\,s}$ in organic field effect transistors (OFETs) [4,5]. Terahertz (THz) spectroscopy studies showed that photoexcitation in DNTT generates free charges, with mobility increasing as temperature decreases, demonstrating a hallmark of the so-called band-like transport behaviour. Modelling of the frequency-resolved THz photoconductivity spectra using the transient localisation theory (TLT) revealed that thermal vibrations dynamically delocalize charges in DNTT along the herringbone plane [6]. The robust molecular design, together with its optimal crystal packing that ensures strong in-plane 2D $\pi$-hole transport, has established DNTT as a leading material for OFET applications. Complementary to this, Takeya's group investigated the excited-state dynamics of DNTT using transient absorption spectroscopy [7]. The study showed that holes and electrons delocalise onto separate molecules to form charge-transfer states, while the remaining neutral Frenkel excitons stay localised on the same molecule.

Building upon these fundamental charge and energy transport studies, DNTT has also been integrated into phototransistors to assess its potential for real-time optoelectronic applications [8–12]. Recent DNTT-based phototransistors exhibited high performance metrics, including photosensitivity, photoresponsivity, and detectivity [12]. For instance, Hung et al. developed an artificial retina utilising DNTT and its alkylated derivatives, achieving a broadband light response with low power consumption[13]. Three main factors dictate the electro-optical properties of a semiconductor: molecular design, crystal packing, and the solid-state morphology. Recent studies have shown that DNTT films grown on $SiO_2$ may display, alongside the prevailing edge-on configuration, a fraction of flat-lying molecular domains which correlate with reduced charge-transport performance, despite the absence of any new crystalline polymorph [14]. Crystal packing, in particular, determines the critical structure-property relationships that enable the rational design of new materials with tailored functionalities. Directing the assembly of molecules into specific crystalline structures with tailored functionalities is a central aim of the rapidly evolving field of crystal engineering. In this regard, the development and discovery of processes to obtain new molecular crystals with two or more distinct structural arrangements, known as polymorphs [15], represent powerful tools to tailor their charge-transport and optoelectronic properties [15–19] .

Polymorphism arises from the existence of multiple local minima in the crystal free-energy landscape, corresponding to distinct molecular packing arrangements with different thermodynamic and kinetic stabilities. At a given temperature, typically only one polymorph is thermodynamically stable and minimises the free energy; the existence of other polymorphs, classified as metastable, arises solely from kinetic barriers that impede transformation into the thermodynamic ground state. Consequently, the thermodynamically stable polymorph may in some cases remain hidden, particularly when its formation requires specific processing conditions and is discovered only later [20]. This phenomenon, known as *late-appearing polymorphism*, is well-documented in the pharmaceutical industry and can have dramatic consequences [21].

Only one reported crystalline structure of DNTT (**Fig. 1a,b**) has already been deposited in the Cambridge Structural Database (CSD refcode: NICLAN) [3]. For low-molecular-weight organic semiconductors, several works distinguish between the thin-film (or substrate-induced) phase and the bulk phase[22–24]. The thin-film phase forms immediately in close proximity to the substrate and typically has a thickness of less than 50 nm [25–27]. Above this layer, a different phase usually develops. Bulk and thin-film phases are usually distinguished by the interlayer (001) spacing[23]. To the best of our knowledge, only one thin-film phase of DNTT has been noted previously, but it has never been structurally resolved[23]. Consequently, the existence of a distinct polymorph of DNTT, as well as its structural, vibrational, and charge-transport properties, has remained unexplored. In this work, we discover, describe, and fully determine the structure of a previously unknown bulk, disordered form of DNTT, which we will refer to as the 'blue' DNTT.

The crystal structure of the *blue* DNTT is shown in **Fig. 1c**. The best-known form of DNTT will be called 'green' DNTT. We distinguish and label the two crystal structures according to their emission colours under UV light. The blue-emitting DNTT phase was first identified when DNTT powder from a Merck bottle was illuminated with a 395 nm UV lamp. While the powder appeared homogeneous, consisting of yellow crystals with similar morphology, under UV illumination, the crystals differed markedly, as shown in **Figs. 1d-e**, indicating the presence of two distinct polymorphs. The UV–Vis absorption spectra of the two DNTT polymorphs — *blue* and *green* DNTT — exhibit similar overall profiles, each characterized by a structured absorption band in the 420–450 nm range (see Fig. 1d), while the emission maximum shifts from 484 to 476 nm for the *green* and *blue* DNTT, respectively (see Fig. 1e and Supplementary Figs. 1, 2).

To elucidate the relative thermodynamic stability of the two polymorphs, we performed slurry experiments, differential scanning calorimetry (DSC), temperature-dependent X-ray diffraction (XRD) measurements, and theoretical calculations. These investigations coherently indicate that the *blue* DNTT is the most stable polymorph at room temperature. XRD measurements reveal an irreversible *green*-to-*blue* phase transition between 210 °C and 240 °C, with the crystal remaining in the *blue* phase upon cooling to room temperature. Moreover, repeated recrystallisation attempts to obtain the *green* DNTT as a bulk powder were unsuccessful, and at present the *green* phase can be isolated in pure form only via thin-film deposition, where the substrate appears to stabilise this metastable phase[14,22]. The *blue* DNTT polymorph exhibits a unique packing motif with interdigitated layers of herringbone-packed molecules, suggesting a more three-dimensional (3D) and isotropic character than the common 2D layered structure of the *green* form. The impact of this distinct crystal packing on the electronic and optical properties is investigated.

By combining XRD, vibrational spectroscopy, optical measurements, and theoretical modelling, we establish how differences in molecular packing govern low-frequency phonons, optical transitions, and, critically, charge-transport dimensionality. THz spectroscopy and TLT calculations reveal that charge transport is highly anisotropic in *green* DNTT, with charge mobility along the electronically active a–b plane and strong suppression along the third crystallographic direction. As a result, *green* DNTT exhibits a predominantly two-dimensional transport network, with holes as the dominant charge carriers.

In stark contrast, the *blue* DNTT exhibits less efficient hole delocalisation within the a–b plane, but a comparable delocalisation along the perpendicular direction, resulting in 3D transport along all crystallographic axes. Notably, in the *blue* polymorph, electrons become the most mobile carriers, with electron mobility along the a- and b-axes exceeding twice the mobility of holes along the same direction in *green* DNTT. To our knowledge, this is the first demonstration of a herringbone-packed organic semiconductor composed of acene-based molecules displaying nearly isotropic, three-dimensional charge delocalisation. The higher-dimensional transport in *blue* DNTT, combined with enhanced electron mobility, highlights the potential of this polymorph to achieve improved charge-transport performance. Although thin films of *blue* DNTT have not yet been realised, our results suggest that stabilising this phase could enable organic semiconductors with superior transport properties and controlled carrier nature compared to the *green* DNTT. More broadly, these findings identify interdigitated herringbone networks as key design parameters for engineering transport dimensionality and high electron mobility in high-performance organic semiconductors.

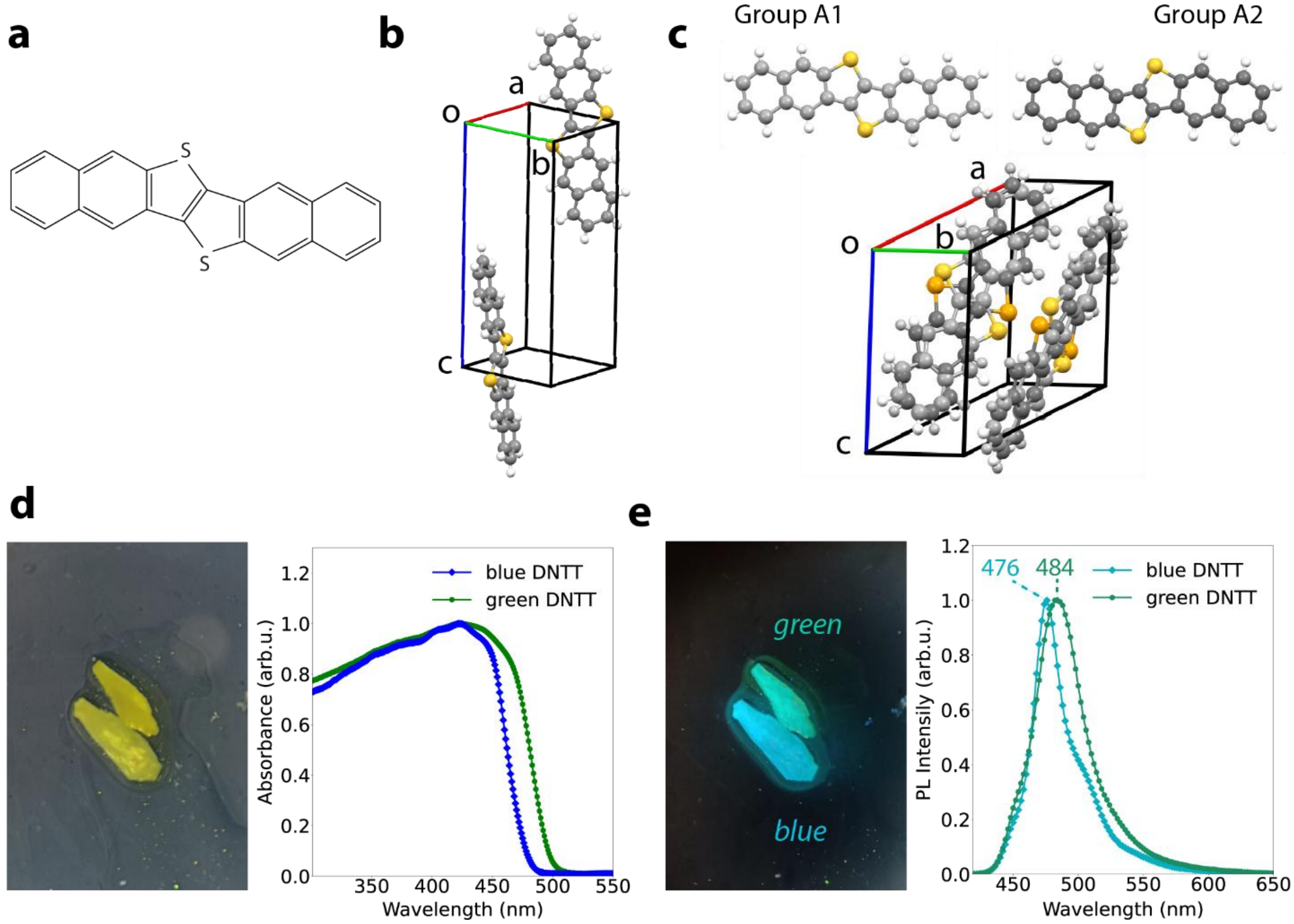

**Figure 1. a,** Molecular structure of DNTT. **b,** Unit cell of *green* **c,** and *blue* DNTT. The crystallographic data are reported in Table 1 and in Supplementary Table 1. The yellow atoms are sulfur atoms, while the grey and white ones are carbon and hydrogen atoms, respectively. The crystal structure exhibits positional disorder, with atoms distributed over two alternative configurations labelled A1 with chemical occupancy around 62% (major form) and A2 with chemical occupancy 38 % (minority form). **d,** Picture of both *blue* and *green* DNTT under visible light, with corresponding UV-vis absorbance spectra measured using *blue* and *green* DNTT pellets (see Methods). **e,** Picture of both *blue* and *green* DNTT under UV (395 nm) light, with corresponding photoluminescence spectra recorded from *blue* and *green* DNTT powders spread between two quartz glasses, excited with 350 nm.

**Results**

**Discovery and confirmation of the *blue* DNTT by XRD measurements**

To confirm and characterise the *blue* DNTT, we prepare higher-quality crystals by dissolving DNTT powder in chloroform and introducing seed crystals of the *blue* DNTT form into the solution[28]. *Blue* DNTT crystallises in the monoclinic space group $P2_1$ (see Table 1 and Supplementary Note 2), with one unique molecule (Z' = 1) per asymmetric unit, as also observed in the *green* polymorph. However, the molecule in the asymmetric unit exhibits positional disorder, corresponding to a 180° rotation about its long molecular axis (see Supplementary Fig. 3), which was modelled with a 60:40 site-occupancy ratio. The persistence of this disorder at 100 K, a temperature where dynamic motion is typically suppressed[29], indicates that it is static in nature (see Supplementary Fig. 4). The disordered packing of the aromatic core is relatively rare among [1]Benzothieno[3,2-b]benzothiophene (BTBT) derivatives: A research on the Cambridge Structural Database (CSD, v6, April 2025[30]) revealed that, out of 198 recorded structures which present the BTBT core[31], only 3 (REFCODEs: DADREJ[32]; KUDDIY and KUDDUK[33]) display a comparable type of disorder.

Both polymorphs exhibit slight deviations from structural planarity (see Supplementary Fig. 5). In the case of *green* DNTT, the nearly planar molecules adopt a 2D layered arrangement in the crystal (**Fig. 2a**), often observed in high-performance organic semiconductors. Within each layer, the molecules adopt a characteristic herringbone packing—typical of polyfused aromatic compounds—which gives rise to a well-defined two-dimensional molecular network with layers defined by the a ([100]) and b ([010]) axes (**Fig. 2 a**). The resulting π–orbital overlap along the two directions defining the herringbone plane enables efficient charge transport across the layer.

In *blue* DNTT, a herringbone packing motif is also present (**Fig. 2 b**), although the molecular layers are oriented differently, with the molecules aligned along the [101] direction. Such an orientation of the herringbone layer is uncommon in organic semiconductor crystals.

In contrast to the *green* polymorph, however, in *blue* DNTT adjacent molecules are displaced by approximately half a molecular length (**Fig. 2 c,d)**. Such a noteworthy shift in molecular arrangements enables substantial interactions between molecules in adjacent layers (light and dark blue in **Fig. 2d**), giving rise to a three-dimensional network. Within the herringbone plane in the *blue* DNTT, one set of molecules (light blue) and the second one (dark blue) are offset by half a molecule, forming interdigitated planes in a three-dimensional packing (**Fig. 2b**). Such a structural reorganization suggests a markedly different landscape for both charge-transport pathways and optical properties, as discussed below (see also Supplementary Fig. 2).

**Table 1.** Crystal data and structure refinement for DNTT polymorphs.

| | ***green* DNTT** [3] | ***blue* DNTT** |
|---|---|---|
| Empirical formula | $C_{22}H_{12}S_2$ | $C_{22}H_{12}S_2$ |
| Formula weight (g mol$^{-1}$) | 340.46 | 340.46 |
| Temperature (K) | 293 | 293 |
| Radiation | MoKα (λ= 0.71073 Å) | MoKα (λ= 0.71073 Å) |
| Crystal system | monoclinic | monoclinic |
| Space group | $P2_1$ | $P2_1$ |
| *a* (Å) | 6.187(4) | 10.5216(8) |
| *b* (Å) | 7.662(6) | 7.5637(5) |
| *c* (Å) | 16.21(1) | 10.5593(7) |
| α (°) | 90 | 90 |
| β (°) | 92.49(2) | 111.698(8) |
| ϒ (°) | 90 | 90 |
| Volume (Å$^3$) | 767.7(9) | 780.79(10) |
| Z, Z' | 2, 1 | 2, 1 |
| Goodness-of-fit on $F^2$ | 0.984 | 1.111 |
| R1 [I > 2σ(I)] | 0.0440 | 0.0893 |
| *w*R2 (all data) | 0.0680 | 0.2021 |

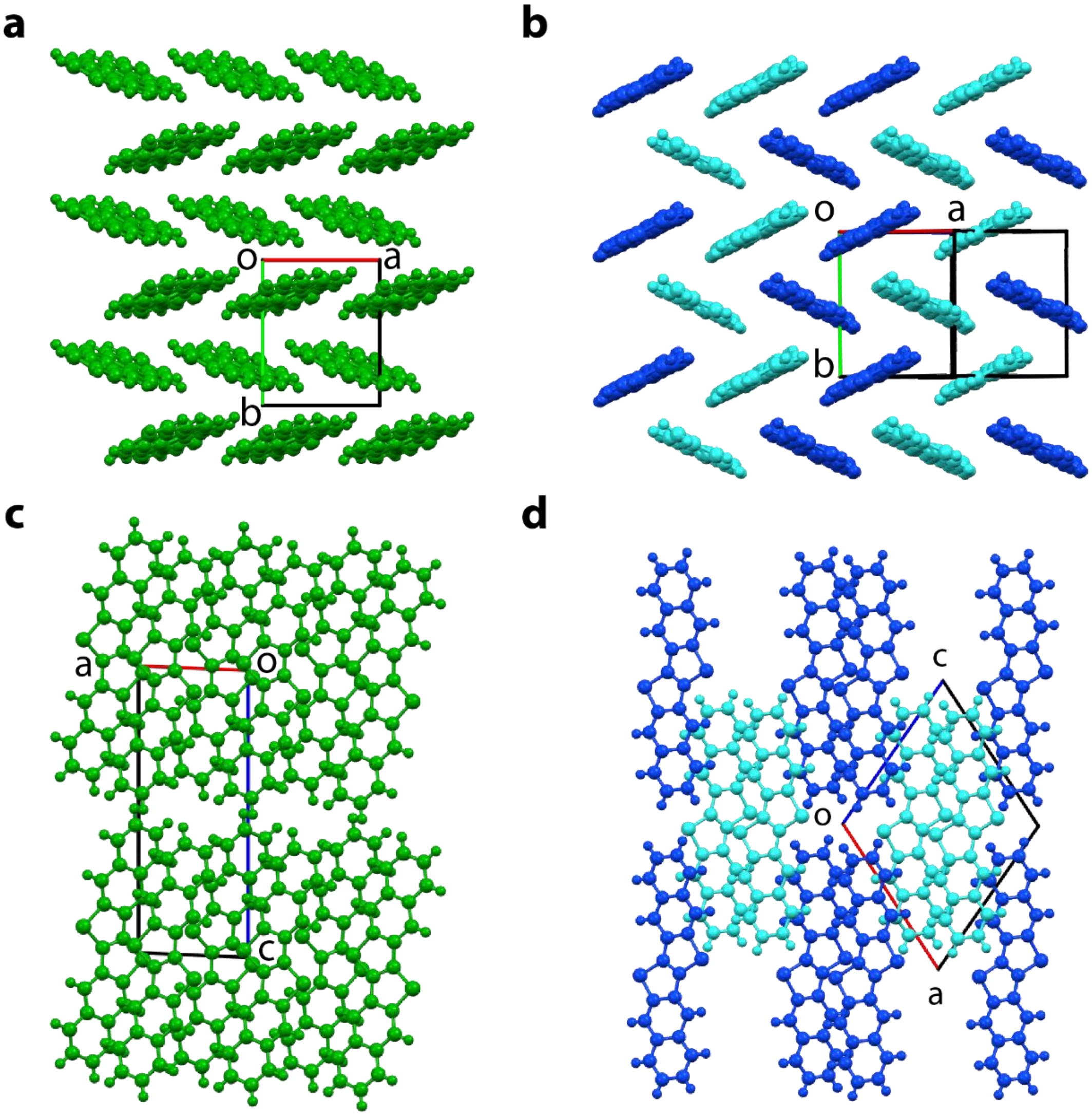

**Figure 2**. Herringbone molecular packing of **a,** *green* and **b,** *blue* DNTT. To highlight the two distinct molecular orientations within the packing, one group of DNTT molecules in the *blue* polymorph is intentionally colored light blue. **c,** Molecular packing view along the b-axis which clearly shows the layers made of the parallel DNTT. **d,** Molecular packing view along the b-axis reveals how the interdigitation of the molecules forms a three-dimensional network. For the sake of clarity, only one of the disordered molecular components (A1) is shown for *blue* DNTT.

**Thermal stability of DNTT polymorphs**

To understand the thermal stabilities of polymorphs, as well as how these structural differences arise, we investigated the temperature-dependent behaviour of the polymorph mixture.

**Figure 3a** shows the XRD patterns of the bare *green* and *blue* DNTT, and a mixed powder containing both phases, alongside patterns calculated from their crystal structures (Supplementary Figs. 6-8). For the mixed powder, the experimental diffraction peaks are well reproduced by a combination of simulated patterns, confirming the presence and identity of the *blue* DNTT phase. We further performed temperature-dependent powder XRD measurements under nitrogen atmosphere using an Anton Paar DHS 900 heating stage. The resulting diffraction patterns are presented in **Figure 3b**. The initial room-temperature measurement reveals that the purchased powder consists of the *green* phase, though clear diffraction peaks from the *blue* phase are also present. Upon heating, a slight but gradual decrease in the intensity of diffraction peaks associated with the *green* phase is observed, accompanied by a slight increase in the *blue* phase, visible on the peak at 8.9° (see Supplementary Fig. 9). The results obtained from temperature-dependent XRD suggest that the *green* phase transforms into the *blue* phase.

Beginning at approximately 210 °C, all X-ray peaks assigned to the *green* DNTT phase rapidly decrease in intensity. Upon heating to 240 °C, the main diffraction peak at 5.4° of the *green* phase initially remains detectable, but then rapidly decreases to 0 upon further heating. In this measurement window, the diffraction peaks belonging to the *blue* phase remain unaffected (see Supplementary Fig. 10). This behaviour at 240 °C may indicate that, in addition to the phase transformation (see also Supplementary Figs. 11-14), partial material loss (e.g., sublimation at elevated temperature of the *green* DNTT) could occur during prolonged annealing. Upon cooling back to room temperature, the remaining DNTT stayed entirely in the *blue* phase.

Overall, these results suggest that the *blue* phase has a lower vapor pressure than the *green* one and is therefore more stable, in agreement with the slurry experiment (see Methods). All the experimental evidence suggests that the *green*–*blue* DNTT system behaves as a monotropic system, with the *blue* form corresponding to the thermodynamically stable phase [34]. Furthermore, DFT calculations confirm that the *blue* form is the thermodynamically stable polymorph, with a free energy 75.7 meV/cell lower than that of the *green* form at room temperature. This increased stability is primarily driven by the electronic energy contribution (see Supplementary Table 2).

Slurry experiments confirm that the *blue* DNTT form is thermodynamically stable at room temperature. DSC measurements of *green* DNTT reveal weak exothermic events, consistent with a transformation from the metastable to the stable form; accordingly, the powder recovered after thermal treatment exhibits blue emission. In contrast, the *blue* phase shows no detectable thermal transitions under the same conditions (see Supplementary Figs. 12-14).

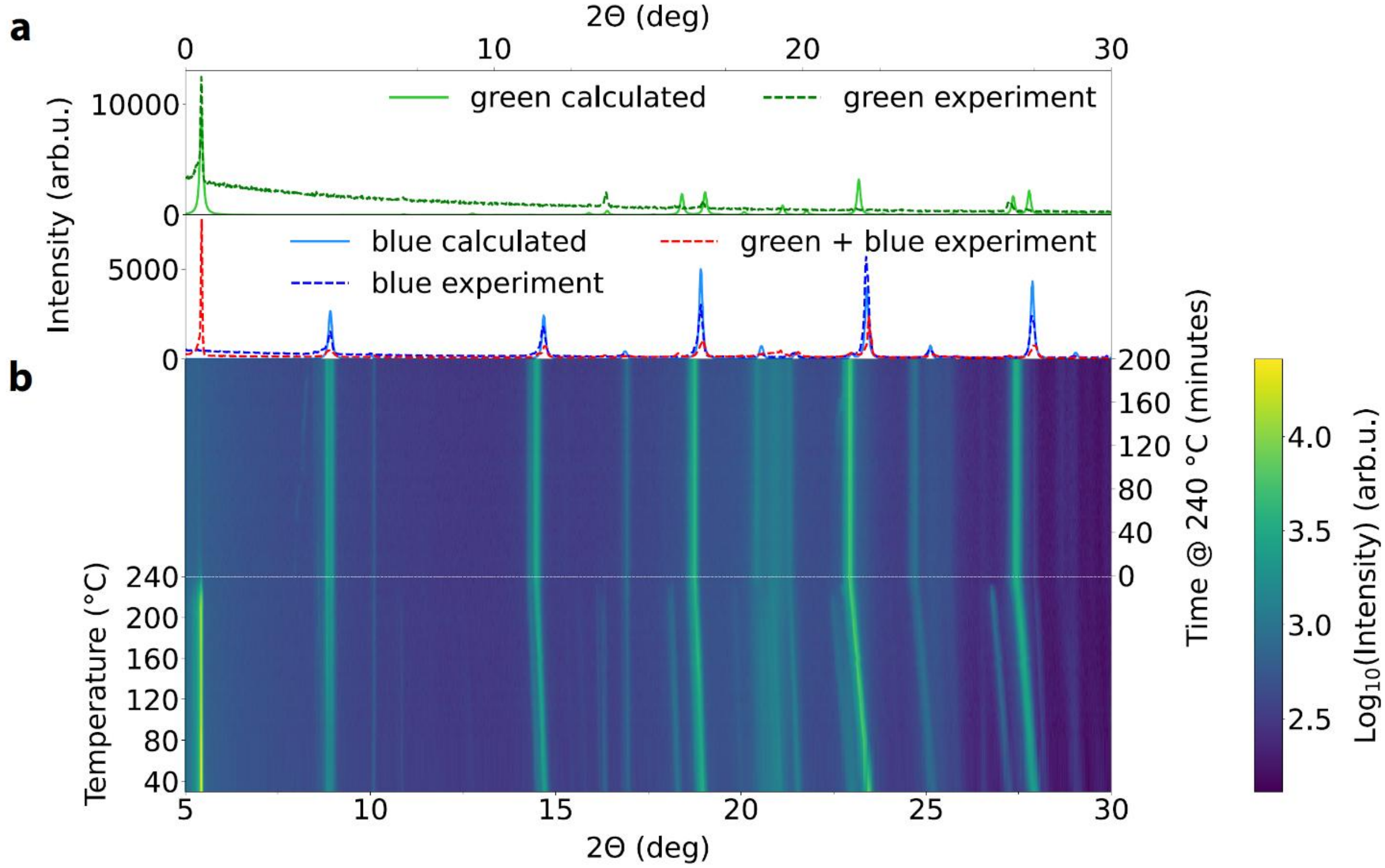

**Figure 3. a,** Experimental and calculated XRD of *green* and *blue* polymorph at room temperature (~25 °C). **b,** Temperature dependence of XRD measurements. The *green* and *blue* phases are present at room temperature. Upon heating above ~120 °C, progressive transformation of the *green* phase into the *blue* phase is shown by the growth of *blue*-related peaks and the simultaneous decline of *green*-associated peaks. Intensities are visualised with logarithmic intensity scaling.

## Raman scattering fingerprints of polymorphism

Further characterisation of the *blue* polymorph was conducted using Raman spectroscopy, enabling a comparative study of the vibrational spectra of the two polymorphs. The crystals of *green* and *blue* DNTT adopt the non-centrosymmetric $P2_1$ space group, which allows vibrational modes to be both IR- and Raman-active. Vibrational spectra of molecular crystals can be heuristically classified into high-frequency (e.g., >300 $cm^{-1}$) modes, which involve the relative motion of atoms within a single molecule, and low-frequency (e.g., <300 $cm^{-1}$) modes, which describe rigid-body intermolecular motions, possibly hybridised with low-frequency intramolecular modes. The low-frequency window of the Raman spectra can be considered as a reliable "fingerprint" for different polymorphs [19,35,36].

Figure 4 shows the Raman spectra of the *blue* and *green* DNTT polymorphs (Fig. 1), comparing experimental and simulated spectra over a broad frequency range. Both polymorphs exhibit several Raman-active vibrations in the low-frequency region (**Fig. 4a**, 10–300 $cm^{-1}$), revealing rich phonon spectra that are remarkably different between the two polymorphs. Solid-state DFT calculations accurately reproduce the low-frequency Raman spectra of both polymorphs, enabling the identification of the prominent vibrational modes. The slightly worse agreement for *blue* DNTT can be attributed to the fact that real crystals present positional disorder while calculations have been performed for the major form (see Supplementary Fig. 3). The experimental Raman spectrum of the *green* DNTT is characterized by a most intense band at 13 $cm^{-1}$, which corresponds to the out-of-phase long-axis sliding of the two molecules in the unit cell (see Supplementary Fig. 15). Such a sliding motion has been reported to be responsible for nearly half of the total transfer integral fluctuations in the *green* DNTT [4] , which are detrimental for charge transport. Interestingly, the sliding motion in the *blue* DNTT is shifted to a higher frequency, around 27 $cm^{-1}$, and presents a significantly weaker intensity than the dominant bands at 57 and 78 $cm^{-1}$, the latter corresponding to mode mixing rigid-molecule motions with intramolecular flexural deformations (see Supplementary Fig. 16).

Complementary time-domain THz spectroscopy in the 13–66 $cm^{-1}$ (0.4-2.0 THz) range has been conducted to further investigate the low-frequency lattice vibrations (see Supplementary Fig. 17 and 18). The long-axis sliding mode is silent in *green* DNTT and appears with modest intensity around 27 $cm^{-1}$ (~0.8 THz) in the *blue* polymorph, in agreement with the infrared absorption calculated at the solid-state DFT level. THz spectroscopy and calculations reveal a prominent absorption feature at approximately 53 $cm^{-1}$, which is nearly identical for both polymorphs. Taken together, Raman and THz vibrational spectra indicate a significantly different vibrational landscape in the low-frequency region for the two DNTT polymorphs, suggesting different effects of thermal lattice vibrations on the energetic disorder experienced by charge carriers and, ultimately, on transport. The excellent overall agreement with experiments provides robust validation of the phonon calculations, supporting their use in electron-phonon coupling analysis and as input for TLT transport simulations.

The high-frequency region of the Raman spectrum (**Fig. 4b**, 300–1700 $cm^{-1}$) is similar for the two polymorphs, confirming the intramolecular nature of these vibrations. The experimental spectra in this spectral window are well reproduced by DFT calculations performed on an isolated DNTT molecule.

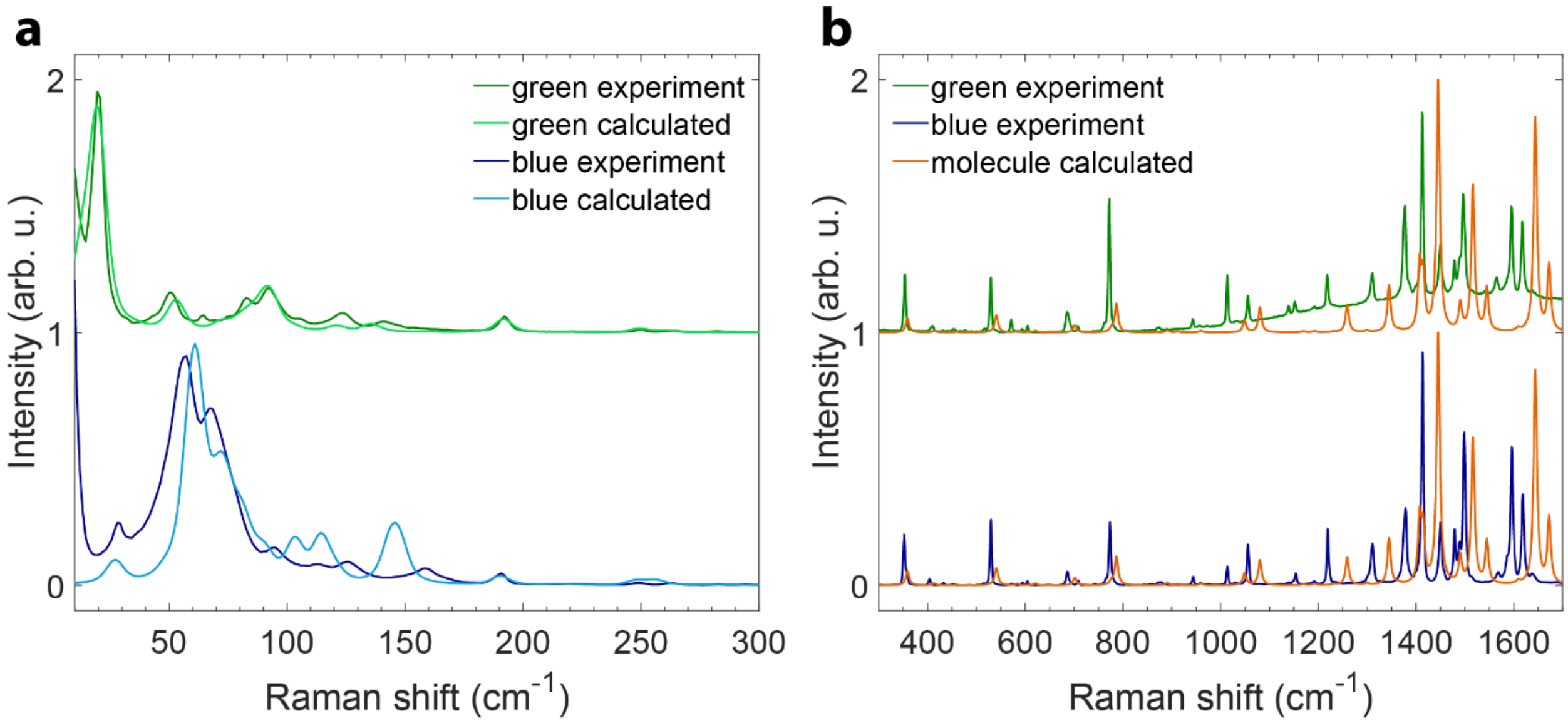

**Figure 4.** Raman spectra of the *green* and *blue* DNTT crystal polymorphs. **a,** The low-frequency region is highly sensitive to the solid-state molecular packing resulting in different spectra for the two forms. Solid-state DFT calculations provide an excellent description of the Raman spectra of the two polymorphs. **b,** The high frequency Raman spectra are very similar for the two polymorphs and can be modelled with isolated-molecule DFT calculations, attesting the intramolecular of the bands.

## Charge transport effects probed by THz spectroscopy

Structural and vibrational spectroscopy data suggest a radically different energy landscape for charge carriers in the *blue* and *green* DNTT. To investigate the charge transport properties of the two polymorphs, we performed optical pump-THz probe (OPTP) measurements. This non-contact optical technique enabled us to photoinject charge carriers and probe photoconductivity and its dynamics without electrodes. We recorded the relative pump-induced change in the THz field, -ΔE/E, normalised to the absorbed photon density ($N_{abs}$). This quantity, measured at the maximum of the THz pulse, is linearly proportional to the real part of the photoconductivity, $\Delta\sigma$ [37]. In the case of diffusive transport, this relationship is expressed as $-\Delta E/E \propto \Delta\sigma = Ne\mu$, where $N$ is the density of photoinduced free charges, $e$ is the elementary charge, and $\mu$ is the charge mobility. In parallel, by fixing the optical sampling at the THz pulse zero-crossing point after the peak field, we can extract the imaginary part of conductivity, which describes the dynamics of both bound and free charges in the semiconductor. As shown in **Fig. 5a**, the peak amplitudes of the real and imaginary parts of the conductivity show little difference between the two structures. On the other hand, the *blue* DNTT's photoconductivity decays faster than that of the *green* polymorph (2.2 ± 0.2 ps vs. 3.2 ± 0.4 ps for the real, and 1.9 ± 0.3 ps vs 4.0 ± 1.1 ps for the imaginary component).

To elucidate the nature and transport properties of charge carriers, we performed a time-domain spectroscopic study at a fixed pump-probe delay time of 1 ps after the photoconductivity maximum. Within the experimental error, the sheet photoconductivity data (see Methods for details) of the two polymorphs are nearly identical at higher frequencies, with slight deviations at low frequencies (see **Fig. 5b**). However, both the charge conductivity dynamics in Fig. 5a and the spectra in Fig. 5b suggest that the imaginary part of conductivity is almost the same, while the real part of the sheet conductivity is slightly higher for the *green* DNTT, in line with the photoconductivity data in Fig. 5a.

The real-part photoconductivity spectra of both the *green* and *blue* polymorphs exhibit an increasing trend with frequency, a feature commonly interpreted as evidence of charge localisation [6,38–41]. In the established framework of the transient localisation theory of charge transport in organic semiconductors, charge transport is limited by dynamic energetic disorder due to thermal molecular motion, causing the localisation of carriers over a length $L$ in a time $\tau_{in}$, related to the vibrational timescale. This theory provides the following analytical formula (see Supplementary Note 8,9) for the charge mobility, $\mu = L^2 e/(2k_B T\tau_{in})$, where $k_B$ is the Boltzmann constant, $e$ is the elementary charge and $T$ is the temperature. Consistent with the transient localization scenario, we analyzed the photoconductivity spectra according to the Drude-Anderson (DA) model [6,38–41] to extract $\tau_{in}$ and the product $NL^2$ as independent fit parameters (see Supplementary Fig. 19). The fitted values for $\tau_{in}$ for the two polymorphs are indistinguishable within the experimental error, namely 265 ± 43 fs and 247 ± 28 fs for the *green* and *blue* DNTT, respectively (Supplementary Fig. 19a). This indicates that, despite the differences evidenced by Raman and THz spectra, the overall impact of the vibrations on the electronic dynamics is similar for the two crystal forms.

**Table 2:** Transient localisation lengths and charge mobilities for the two DNTT polymorphs along the crystallographic axes. The localisation length is obtained from TLT simulations, and the mobility is obtained by combining $L^2$ with $\tau_{in}$ extracted from OPTP measurements (Drude-Anderson model fit). *a* and *b* correspond to the crystallographic axes, $c^*$ is perpendicular to the a-b plane. The $c^*$ components of $L^2$ and $\mu$ for the *green* polymorph are zero according to the strictly 2D model adopted for this layered semiconductor.

| | | **Holes** | | **Electrons** | |
|---|---|---|---|---|---|
| **Polymorph** | **axis** | $L^2$ **(Å²)** | $\mu$ **(cm²/Vs)** | $L^2$ **(Å²)** | $\mu$ **(cm²/Vs)** |
| *Green* DNTT | *a* | 404.3 ± 4.0 | 3.08± 0.50 | 181.0 ± 1.3 | 1.38 ± 0.22 |
| | *b* | 250.2 ± 2.2 | 1.91±0.31 | 187.4 ± 1.2 | 1.43 ± 0.23 |
| | $c^*$ | 0 | 0 | 0 | 0 |
| *Blue* DNTT | *a* | 299.3 ± 2.8 | 2.45 ± 0.28 | 808.5 ± 4.0 | 6.61 ± 0.74 |
| | *b* | 177.8 ± 1.4 | 1.45 ± 0.16 | 499.2 ± 2.2 | 4.08 ± 0.46 |
| | $c^*$ | 138.5 ± 1.3 | 1.13 ± 0.13 | 608.4 ± 3.0 | 4.97 ± 0.56 |

To disentangle the effects of carrier density and charge delocalisation on photoconductivity, and to gain insight into the differences in electronic structure and transport properties between the two polymorphs, we performed TLT simulations based on input from DFT calculations. The analysis of intermolecular transfer integrals confirms that *green* DNTT is a 2D system with charges confined to the a-b molecular plane and negligible inter-plane interactions (see Fig. 2 a,c and Supplementary Table 5), while the *blue* form exhibits a marked 3D character. Such a remarkable feature directly follows from the interdigitated molecular layers that form the distinctive herringbone structure shown in Fig. 2 b,d, characterised by molecules shifted by half their lengths along the long-axis direction. This results in a significant reduction in the transfer integrals between coplanar molecules compared to the *green* form, but it yields to sizable values for molecular pairs offset along the long axis (see Supplementary Table 6).

Transfer integrals, together with thermal fluctuations (see Methods for details), are then used to compute the localisation length within the TLT framework. The resulting $L^2$ values for the major form reported in Table 2 show that *blue* DNTT, unlike the *green* form, exhibits charge delocalisation comparable along the three directions, clearly indicating a 3D nature for the charge transport network determined by the crystal packing. Notably, whereas holes have a larger localisation length than electrons in the *green* DNTT, this relationship is inverted in the *blue* phase, where electrons become more delocalized (see Supplementary Fig. 20 for the electronic band structures of both polymorphs).

The Drude Anderson model applied to THz experiments provides direct access only to $NL^2$ and the scattering time $\tau_{\mathrm{in}}$, however, by combining the experimental $\tau_{in}$ from OPTP measurement and the calculated $L^2$, we can estimate carrier mobilities, given in **Table 2** and **Fig. 5c**. Our analysis reveals that the electron mobility in the *blue* DNTT exceeds the hole mobility in the *green* DNTT by more than a factor of two along the a- and b-axes. Moreover, charge transport in the *green* form is strictly confined to the crystalline a-b plane, while the mobility is distinctly isotropic, being nearly equivalent along all three crystalline directions in the *blue* DNTT. Enabled by its unique herringbone-interdigitated structure, *blue* DNTT represents a rare case where the dimensionality of charge transport is fundamentally enhanced, moving beyond traditional 2D constraints (see **Figs. 5d, e**). Although thin films of the *blue* phase are not yet available, its 3D transport character and enhanced electron mobility along the a-axis could be directly probed in single-crystal OFET devices through orientation-resolved measurements. The additional conduction channel along the previously inactive axis may contribute to a more isotropic charge transport landscape, suggesting a practical advantage of the *blue* polymorph for efficient charge separation in bulk or thicker films, opening avenues for next-generation high-performance organic electronic devices.

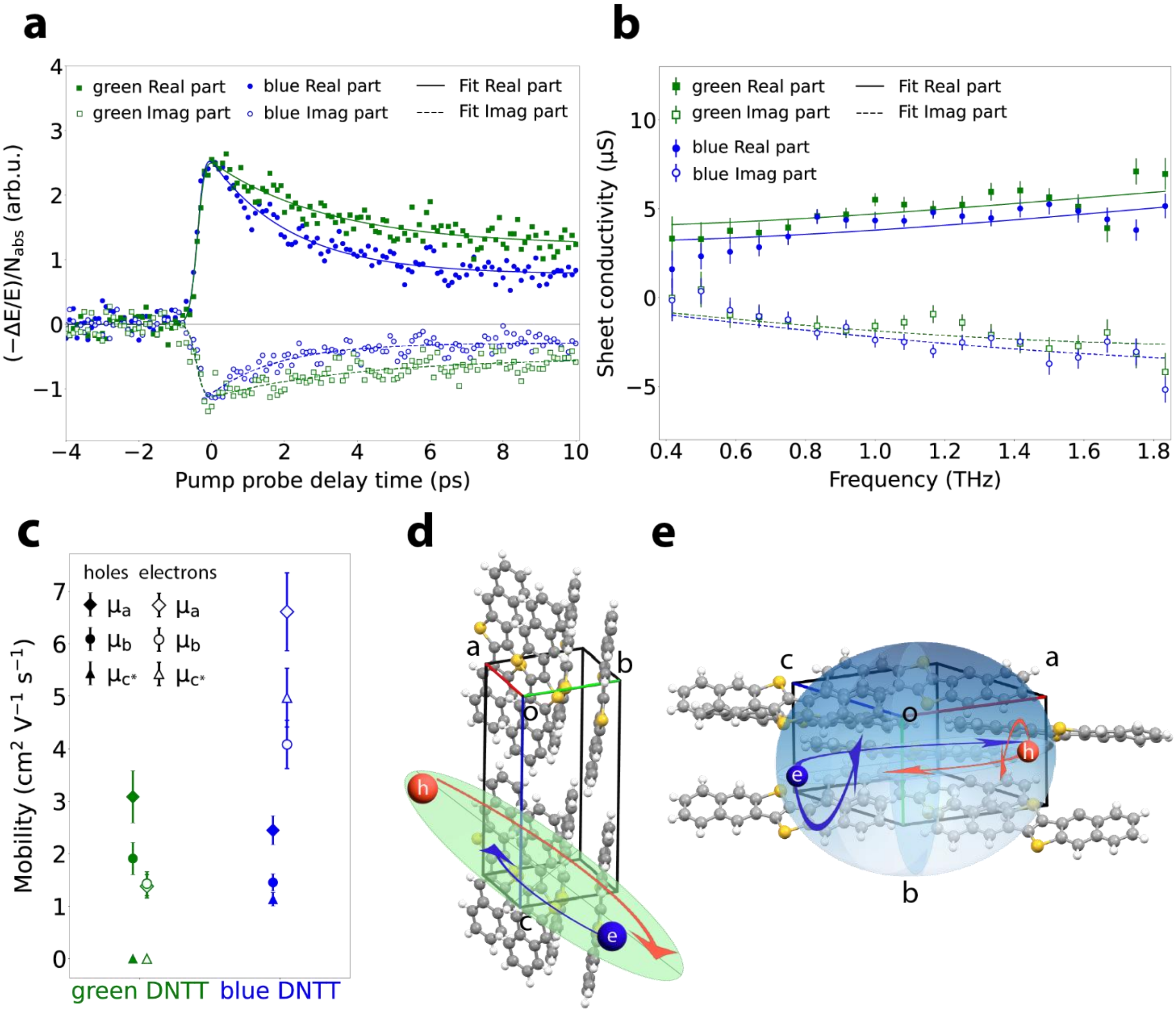

**Figure 4: a,** Relative change $-\Delta E/E$ (normalized to the absorbed photon density $N_{abs}$) of the THz electric field for the *blue* and *green* pellets. The filled markers indicate the real part, and the open markers indicate the imaginary part. The lines represent exponential fits. **b,** Frequency-dependent THz spectra of the two pellets recorded 1 ps after excitation with 50 fs pulses with a photon energy of 3.1 eV. The lines are a fit to the DA model described in Supplementary Note 8. Solid and dashed lines refer to the real and imaginary parts of the photoconductivity, respectively. **c,** Mobilities extracted using the TLT model. $\tau_{in}$ is the average extracted from the DA model applied on three separate experiments and $L^2$ is obtained by theory. The *green* phase shows two-dimensional (a-b plane) transport, whereas the *blue* phase displays three-dimensional transport (c* is referred to the direction perpendicular to the a-b plane). **d,e,** Anisotropy of charge delocalisation: in the *blue* DNTT, the blue ellipsoid illustrates how holes (h) and electrons (e) are spatially delocalized (along a length *L*), with longer axes indicating greater delocalisation along the crystallographic directions. In *green* DNTT, the ellipsoid collapses into a green ellipse in the a–b plane, showing that charge delocalisation is confined exclusively to this plane.

## Conclusions

We report on the discovery of the new polymorph of DNTT, one of the most air-stable and conductive organic semiconductors, which has been extensively explored for electronic and photonic applications. This newly identified form, termed the *blue* DNTT, can be readily distinguished by its blue-shifted emission relative to the green emission of the well-known polymorph, now named the *green* DNTT. Notably, the *blue* DNTT is the thermodynamically stable form and is already present, albeit unnoticed, in commercial DNTT powders; it also arises through a thermally driven transformation of the well-known *green* form. The *green*- *blue* DNTT system is likely monotropic, with the metastable *green* phase becoming increasingly difficult to isolate in bulk, as trace nuclei of the stable *blue* polymorph preferentially direct crystallisation toward the *blue* form (the "universal seeding" effect)[42,43]. In contrast, the *green* DNTT phase appears to form preferentially during thin-film deposition.

By combining THz spectroscopy and TLT simulations, we show that the *blue* DNTT herringbone-interdigitated structure reshapes both the transport dimensionality and the relative mobility of charge carriers. In the *green* polymorph, charge transport is predominantly two-dimensional and dominated by holes, with higher mobility along the electronically active directions arising from a larger in-plane localisation length.

The scattering time $\tau_{in}$ remains essentially constant, suggesting that differences in transport primarily originate from changes in electronic coupling rather than vibrational dynamics.

The *blue* polymorph exhibits higher-dimensional transport pathways, enabling charge propagation along all three crystallographic directions. Remarkably, electrons become the fastest carriers in the *blue* phase, with mobility along the a- and b-axes exceeding twice that of holes along the same axes in *green* DNTT. To the best of our knowledge, our results establish 3D charge transport in acene-based semiconductors for the first time, demonstrating that herringbone-packed molecular solids can sustain isotropic charge delocalisation. Not only does *blue* DNTT overcome reduced in-plane delocalisation compared to the *green* phase, but it also shows enhanced electron mobility reaching ~7 $cm^2$/V s along the a-axis, more than twice the hole mobility in the *green* phase. While thin films of *blue* DNTT have not yet been realized, our results suggest that stabilising this polymorph could enable organic semiconductors with improved electron transport performance.

Beyond identifying a readily processable *blue* emitter, our discovery also highlights the importance of polymorphism in controlling both the dimensionality and the carrier type in charge transport, opening new directions for the design of future high-performance DNTT-based optoelectronics.

***Material and methods***

DNTT was purchased from Merck and used for the experiments without further purification. The selection of crystals corresponding to the *blue* and *green* phases was carried out manually using a UV torch to trigger luminescence. The selected crystals were then ground into powder (when necessary) and analyzed using a powder diffractometer to ensure that only the desired phase was obtained.

***Recrystallisation from solution.*** 0.5 mmol of DNTT (0.170 g), sourced from a commercial vial containing both polymorphs, was dissolved in chloroform or dichloromethane in a beaker. The solvent was then evaporated, resulting in the precipitation of a yellow powder. The crystalline powder was subsequently analyzed using powder X-ray diffraction, which confirmed the exclusive presence of the *blue* DNTT polymorph. Several attempts have been made to obtain pure *green* form by fast recrystallisation, but in all cases, *blue* DNTT or a mixture of *green* and *blue* DNTT was obtained. Pure *green* DNTT can be obtained by thin film deposition.

***Recrystallisation for crystals of blue form.*** To obtain higher-quality crystals of the *blue* polymorph, 0.5 mmol of commercial DNTT (0.170 g) was dissolved in chloroform in a beaker. The resulting solution was transferred into small test tubes and covered with perforated Parafilm to allow for slow solvent evaporation. In some tubes, seed crystals of the *blue* DNTT polymorph were introduced to promote selective nucleation and growth. This procedure yielded single crystals suitable for SCXRD analysis. Data collection was also attempted using commercially available *blue* DNTT crystals; however, these were frequently twinned, which complicated structure refinement and resulted in lower-quality diffraction data.

***Conversion procedures***

***Slurry.*** 0.5 mmol of DNTT (0.170 g), sourced from a commercial vial containing both polymorphs, was placed in a glass vial with 1 mL of solvent (acetonitrile, chloroform, or dichloromethane) and stirred in a glass vial for 24 hours. When acetonitrile was used as the solvent, the stirring duration was extended to 4 days. The resulting crystalline powders were analyzed via powder X-ray diffraction, revealing the exclusive presence of the *blue* DNTT polymorph.

***Thermal conversion*****.** The thermal conversion was observed by hot stage microscopy. Crystals of *green* DNTT were heated at 10°C $min^{-1}$; the conversion into *blue* DNTT was observed at around 250°C. The conversion has been confirmed via powder X-ray diffraction, revealing the exclusive presence of the *blue* DNTT polymorph.

***Pellet preparation.*** Pellets were prepared by pressing a mixture of the desired DNTT phase powder with Poly(tetrafluoroethylene)(PTFE—a polymer transparent to terahertz radiation). *Blue* DNTT pellet: 20 mg of *blue* DNTT powder and 428.37 mg of PTFE. *Green* DNTT pellet: 20 mg of *green* DNTT powder and 431.04 mg of PTFE. Pressing was performed at 7 tons for 4 minutes. The pellet containing the *green* DNTT used for the measurements was 0.95 ± 0.02mm thick, while the one containing the *blue* DNTT was 0.86 ± 0.02 mm thick. Crystallite sizes were estimated using the modified Scherrer equation [44], based on the powder X-ray diffraction patterns shown in Supplementary Fig. 8. Only reflections assigned to the DNTT polymorphs were considered, excluding peaks associated with PTFE. The analysis revealed that *green* DNTT crystallites dispersed in PTFE had an average size of 29.1 ± 0.1 nm, while *blue* DNTT crystallites exhibited a slightly smaller average size of 26.0 ± 0.1 nm.

***Structure determination by X-ray diffraction***

The single crystal data of *blue* DNTT were collected at room temperature on an Oxford Xcalibur using Mo-Kα radiation, equipped with a graphite monochromator and CCD Sapphire detector. The single crystal data of *blue* DNTT were collected at 100 K on a Bruker D8 Venture using a microfocus Mo-Kα source, equipped with Photon III detector. Crystal data details are summarized in Supplementary Table 1. The SHELXT [45]and SHELXL[46] algorithms were used for the solution and refinement of the structures based on $F^2$. All the atoms, excluding the hydrogens, were refined anisotropically. Hydrogen atoms have been added to the theoretical positions. CCDC 2476561-(RT) and 2497256 (100K) contain the supplementary crystallographic data for this paper. The data can be obtained free of charge from The Cambridge Crystallographic Data Centre via www.ccdc.cam.ac.uk/structures. The graphical representations of the structures were displayed with the software MERCURY[47].

***Hotstage microscopy (HSM)***

The analysis was performed using an OLYMPUS BX41 microscope equipped with a VISICAM 5.0 camera and a system for temperature control, Linkam TMS 94. The photos were taken under polarized light to underline the modification due to the solid-state transition, with a 10× magnification.

***DSC***

Calorimetric measurements were performed with a DSC Q10 TA instrument equipped with an RCS intracooler. The samples in aluminum open pans were heated at 10° C/min. The results are presented and discussed in the Supporting Information.

***TGA***

Thermogravimetric analyses were performed using a Perkin-Elmer TGA8000. The measurements were carried out under nitrogen flow with a 10° C/min gradient. The results are presented and discussed in the Supporting Information.

***Raman Spectroscopy***

The measurements were performed using a WiTec alpha300 Raman spectrometer equipped with a 785 nm laser.

***UV-Vis Spectroscopy***

The measurements of pellets were performed in a UV/VIS/NIR spectrometer Lambda 1050+ from PerkinElmer. To better quantify the light absorption of the two DNTT-based pellets, we placed the samples in an integrated sphere. We used as a reference a PTFE pellet obtained with the same procedure described in pellet preparation and with a resulting thickness 0.89 ± 0.02 mm.

***Photoluminescence Spectroscopy***

The emission spectra of the compounds in the solid state have been collected in front-face mode with an Edinburgh FLS920 fluorimeter equipped with a Peltier-cooled Hamamatsu R928 PMT (280−850 nm), and corrected for the wavelength-dependent phototube response.

***Temperature-dependent powder XRD***

Temperature-dependent XRD experiments were conducted using a Panalytical Empyrean powder diffractometer equipped with an Anton Paar DHS 900 heating stage using Cu-Kα radiation. The purchased DNTT powder was ground using a mortar and pestle, deposited onto a silicon wafer and placed on the heating stage. The powder was subsequently enclosed under a polymer dome and maintained under a nitrogen atmosphere. After an initial powder XRD measurement at room temperature (26 °C), the powder was heated to 80 °C at a heating rate of 5 °C/min. Following a 10 min equilibration time at this temperature, heating cycles were performed between 80 °C and 240 °C in 5 °C increments. At each temperature step, an equilibration time of 8 minutes was used to ensure uniform sample temperature, followed by an XRD measurement with an acquisition time of 4 min. After reaching 240 °C, continuous measurements were performed to check the thermal stability of the obtained DNTT powder.

***Density functional theory calculations of vibrational properties***

Phonons have been computed with periodic DFT calculations for the two phases of DNTT adopting the Minimal Molecular Displacements method implemented in the PhonoMol code [48]. DFT calculations employed the Perdew−Burke−Ernzerhof (PBE) functional [49]together with the projector-augmented wave method (PAW, version 6.4) and Grimme D3 dispersion corrections with Becke−Johnson damping (D3-BJ)[50,51], using the Vienna Ab initio simulation package (VASP, version 6.4.0)[52–55]. The input parameters of DFT calculations (see Supplementary Table 7) have been chosen to ensure the convergence of the total energy within 1 meV/atom, as required for accurate phonon calculations in the THz region. The normal modes of the isolated DNTT molecule, required to set up the basis of molecular displacements, have been computed at the PBE/def2-SVP level including D3-BJ dispersion corrections with the ORCA code [56,57]. A cutoff frequency of 500 $cm^{-1}$ has been employed within the Minimal Molecular Displacements method. Force constant matrices have been computed from supercell calculations using 2×2×2 crystal supercells. Vibrational free energies have been computed for an 8x8x8 sampling mesh for the Brillouin zone, ensuring converged results. Isolated molecule Raman spectra have been calculated at the B3LYP/6-31G(d) level with the Gaussian 16 simulation package. Raman spectra have been calculated as a sum of Lorentzian peaks (HWHM 4 $cm^{-1}$), with nonresonant intensities accounting for experimental conditions (laser wavelength 785 nm, T=300 K)[58].

***Energetic disorder and charge mobility calculations***

Localization length and charge mobilities have been computed within the transient localization theory, adopting the relaxation time approximation and exact diagonalizations [59]. Transport simulation results in the main text have been obtained for a typical value of $\hbar/\tau_{in} = 2.5\ \mathrm{meV}$ – see Table Supplementary Table 8,9 for an analysis of the effect of this parameter. Large crystal supercells (up to 64x32x1 and 20x10x10, respectively, for *green* and *blue* DNTT, ensuring converged results) of each polymorph have been mapped to tight-binding models for hole particles, applying periodic boundary conditions. The lattice sites coincide with individual molecules (HOMO site orbital). Transfer integrals (J) between neighboring molecules have been evaluated at the PBE/def2-SVP level with the dimer projection method[60]. Only intermolecular pairs with hole transfer integral above 3 meV have been included in the tight binding model, resulting in a 2D- and 3D-periodic models for the *green* and *blue* phases, respectively. The gradients of J with respect to Cartesian atomic coordinates have been computed at the PBE/def2-SVP level, enabling the calculation of the thermal energetic disorder within the framework of a linear nonlocal electron-phonon coupling model [4,61]. Thermal vibrations are described within classical statistics, and the temperature is set to 300 K. The variance of the transfer integral has been computed with an 8x8x8 Brillouin-zone sampling, ensuring converged results against phonon band dispersion.

***Optical Pump THz Probe Spectroscopy***

The THz setup is described in ref. [62]. We used a Ti:sapphire laser having 800 nm central wavelength, duration 50 fs and repetition rate of 1 kHz. The THz field is generated by optical rectification in a Zinc telluride (ZnTe) crystal cut along the 110 direction. The pellets were excited by a 400 nm pulse, which was generated by second harmonic generation of the Ti:sapphire laser in a beta barium borate (BBO) crystal. We measure the change in the THz transmission induced by photoinjected charges, by monitoring the peak of the THz field $\Delta E = E_p - E$, where $E_p$ is the transmitted THz field through the pumped pellet and $E$ through the unpumped pellet. The charge dynamics in Fig. 5a reports the ratio between $\Delta E$ and $E$, normalized with the absorbed photon density $N_{abs}$. $N_{abs}$ is calculated by knowing the incident photon density $\varphi_{\mathrm{i}}$( and the absorbance (A)) around 400 nm of the two pellets , using $\varphi_{\mathrm{i}}(1 - 10^{-\mathrm{A}})$. $\Delta E/E$ is proportional to the sheet conductivity $\Delta\sigma_s$:

$$\Delta\sigma_s = \left(-\frac{\epsilon_0 c\ (n_{\mathrm{u}} + n_{air})}{E} \cdot \Delta E\right)$$

Where $n_{air}$ (~1) the refractive index of the air and $n_u$is the THz refractive index of the unpumped pellet, set to 1.5 as the refractive index of PTFE as reported in [63] and also measured by us and reported in Supplementary Fig. 18.

***Contributions***

L.D.V. and C.Z. designed the research project. L.D.V. performed all the Raman, UV-Vis and THz measurements analyzing the data using DA model. G.W. contributed through scientific discussions on the THz results. C.Z. performed the photoluminescence measurements, HSM, conversion measurements, the XRD measurements at room temperature and solved the *blue* DNTT structure. C.Z and L.D.V. found the polymorph with UV light. M. Bardini performed all solid-state vibrational calculations and TLT calculations to estimate $L^2$ for mobility with assistance from G.D. L.S. is the primary developer for Phonomol, coded all the features used by M. Bardini and discussed with G.D, M. Bardini and L.D.V. M.S. and R.C.-H. performed and analyzed Raman and UV-Vis calculations with the Gaussian package. F.G. conducted temperature-dependent XRD measurements proposed by A.G. Data were analyzed by A.G., C.Z. and F.G. J.J.G. performed preliminary UV-Vis measurements and contributed to discussions on the THz experiments. L.D.V, C.Z, G.D, H.I.W., M. Bonn and L.M. wrote the manuscript. All authors reviewed and discussed the manuscript.

***Acknowledgment***

L.D.V. would like to acknowledge Samir Al-Hilfi and Peigen Yao for useful discussions. L.D.V is grateful to Marc-Jan van Zadel and Florian Gericke for their excellent technical support, in particular for the design and fabrication of the pellet sample holders used for the UV-Vis and THz spectroscopy measurements. We express our gratitude to Guillaume Schweicher for valuable scientific discussions. Z.C and L.M. are grateful to Davide Balestri for his valuable assistance in the collection of diffraction data and in the determination and refinement of the DNTT crystal structure under low-temperature conditions. J.J.G. gratefully acknowledges financial support from the Alexander von Humboldt Foundation.